\documentclass[conference]{IEEEtran}
\IEEEoverridecommandlockouts
\usepackage{cite}
\usepackage{amsmath,amssymb,amsfonts}
\usepackage{multirow}
\usepackage{physics}

\usepackage{algorithmic}
\usepackage{graphicx}
\usepackage{textcomp}
\usepackage{xcolor}

\usepackage{graphicx}  

\usepackage{hyperref}  
\usepackage{multirow}
\usepackage{hyperref}  
\usepackage{amsmath}   
\usepackage{amsfonts}  

\hypersetup{
  colorlinks=true,  
  urlcolor=blue,    
  linkcolor=blue    
}

\def\BibTeX{{\rm B\kern-.05em{\sc i\kern-.025em b}\kern-.08em
    T\kern-.1667em\lower.7ex\hbox{E}\kern-.125emX}}
\begin{document}


\title{PhishVQC: Optimizing Phishing URL Detection with Correlation Based Feature Selection and Variational Quantum Classifier}

\author{\IEEEauthorblockN{Md. Farhan Shahriyar}
\IEEEauthorblockA{\textit{Dept. of Computer Science and} \\
\textit{Engineering} \\
\textit{World University of Bangladesh} \\
Dhaka, Bangladesh \\
farhanshahriyar.cse1@gmail.com}
\and
\IEEEauthorblockN{Gazi Tanbhir}
\IEEEauthorblockA{\textit{Dept. of Computer Science and} \\
\textit{Engineering} \\
\textit{World University of Bangladesh} \\
Dhaka, Bangladesh \\
gazitanbhir@gmail.com}
\and
\IEEEauthorblockN{Abdullah Md Raihan Chy}
\IEEEauthorblockA{\textit{Dept. of Computer Science and} \\
\textit{Engineering} \\
\textit{Southern University Bangladesh} \\
Chittagong, Bangladesh \\
raihanchowdhuryengineer@gmail.com}
\and
\IEEEauthorblockN{Mohammed Abdul Al Arafat Tanzin}
\IEEEauthorblockA{\textit{Dept. of Computer Science and} \\
\textit{Engineering} \\
\textit{Brac University} \\
Dhaka, Bangladesh \\
tanzinabdul@gmail.com}
\and
\IEEEauthorblockN{Md. Jisan Mashrafi}
\IEEEauthorblockA{\textit{Dept. of Computer Science and} \\
\textit{Engineering} \\
\textit{Brac University} \\
Dhaka, Bangladesh \\
jisanmashrafi542@gmail.com}

}

\maketitle

\begin{abstract}
Phishing URL detection is crucial in cybersecurity as malicious websites disguise themselves to steal sensitive information. Traditional machine learning techniques struggle to perform well in complex real-world scenarios due to large datasets and intricate patterns. Motivated by quantum computing, this paper proposes using Variational Quantum Classifiers (VQC) to enhance phishing URL detection. We present PhishVQC, a quantum model that combines quantum feature maps and variational ansatzes such as RealAmplitude and EfficientSU2. The model is evaluated across two experimental setups with varying dataset sizes and feature map repetitions. PhishVQC achieves a maximum macro average F1-score of 0.89, showing a 22\% improvement over prior studies. This highlights the potential of quantum machine learning to improve phishing detection accuracy. The study also notes computational challenges, with execution wall times increasing as dataset size grows.
\end{abstract}


\section{Introduction}
Phishing attacks typically begin with a redirection or manipulation of the victim into clicking a malicious link leading to efforts aimed at exfiltrating sensitive information such as login credentials or financial details. As phishing techniques evolve over time  the threat landscape in cyberspace becomes increasingly sophisticated \cite{r1}
\cite{r2}.

This appears more challenging to effectively counter such attacks. While classical machine learning approaches have made significant progress in detecting phishing attempts the rapidly evolving nature of phishing necessitates more advanced solutions.

\begin{figure}[htbp]
    \centering
    \includegraphics[width=\linewidth]{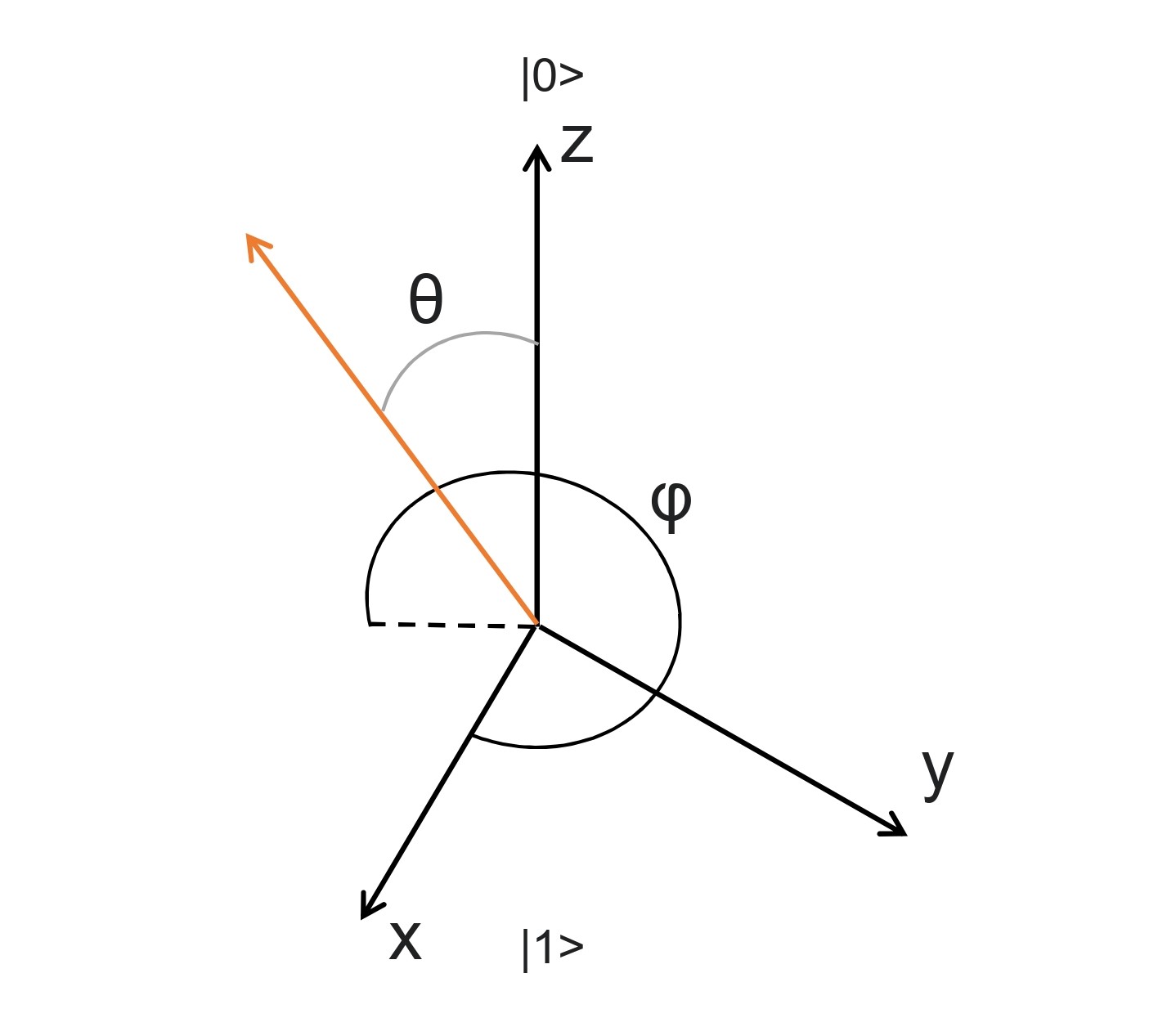}  
    \caption{Visual representation of the Bloch Sphere , the state of a single Qubit.}
    \label{fig:qc}
\end{figure}

Quantum machine learning (QML) presents a promising solution to these challenges given its ability to process large computations in a significantly shorter time frame. QML is powered by Quantum Computing extending its area on machine learning. Quantum computers process data through qubit superposition.In fig \ref{fig:qc} it is represented by the yellow vector on

the Bloch Sphere. Unlike classical bits (0 or 1), qubits exist as a combination of $\ket{0}$  and $\ket{1}$ , defined by angles \( \theta \) and \( \phi \) . 

The  qubits are being used  to encode classical data into methods like amplitude or angle encoding
where features determine the qubit's state. Quantum gates manipulate these states in parallel,
enabling efficient computations, such as processing phishing dataset features (f1,f2...fn )
simultaneously. However, challenges like inefficient encoding, limited qubits, gate errors, and
decoherence hinder progress. Optimized algorithms, better encoding, and advanced hardware are
crucial for QC’s success.

However, its application is currently limited to certain fields. QML also demonstrates a superior capacity for identifying complex patterns compared to classical methods. Recently, the use of quantum machine learning in cybersecurity has emerged as a new frontier\cite{r3}. Quantum computing is opening new possibilities for more powerful computational capabilities\cite{r4}.
Although much of the current research on QML cybersecurity focuses on fraud detection through feature encoding and variational quantum circuits, its application to phishing URL detection remains an underexplored area\cite{r5}\cite{r6}. Given the growing sophistication of phishing techniques there is an urgent need to explore quantum based solutions for effective phishing detection. This presents a critical opportunity for advancing the field of phishing detection through quantum computing.

Previous studies in quantum machine learning have primarily relied on small datasets  (e.g., 200 data points) \cite{r10} and there has been limited research on how the scalability of data points affects the quantum classification process. Additionally, most classical machine learning models have incorporated large sets of features, yet their integration with quantum processing has not been thoroughly investigated. This study aims to address these gaps by focusing on the following key contributions:

\begin{itemize}
    \item Analyzed how the increase in the number of data points affects the performance of quantum algorithms in phishing URL detection.
\end{itemize}

\begin{itemize}
    \item  Examined the integration of correlation based classical feature selection with quantum models to improve detection accuracy in phishing URL classification.
\end{itemize}

\begin{itemize}
    \item Introduced PhishVQC a novel quantum model specifically designed to classify phishing URLs more efficiently.
\end{itemize}

This paper presents a literature review in Section II and details the methodology used in Section III. Section IV covers the experiments and results while Section V concludes with discussions and future work.

\section{LITERATURE REVIEW}
The study \cite{r8} demonstrated the impact of correlation independency on feature selection and algorithm performance highlighting its significance in enhancing the results of machine learning models.The paper by \cite{r9} explores a hybrid classical quantum approach that integrates classical and quantum algorithms in an ensemble model to improve fraud prevention decisions. In this context the Quantum Support Vector Machine (QSVM) offers a complementary exploration of the feature space resulting in improved accuracy for the mixed quantum - classical model in fraud detection when applied to a significantly reduced dataset. Notably, the new algorithm identified features with minimal overlap in the correlation matrix suggesting that these feature choices are viable for improving the detection process.

Reyes-Dorta et al.\cite{r10} developed a quantum machine learning approach for detecting malicious URLs using a Variational Quantum Classifier (VQC). The dataset consisted of 145,532 random data points, including 6,965 fraudulent URLs, from which 200 observations were selected for analysis. To map the data into quantum states, the study tested several ansatzes, including RealAmplitudes (RealAmp), EfficientU2, and ExcitingPreserving (ExcitPreserving). Optimizers such as COBYLA, GradientDescent, and SLSQP, each running for 20 iterations, were also evaluated. The results showed that the combination of the RealAmplitudes ansatz (with two repetitions) and the SLSQP optimizer produced the best performance. The VQC model achieved a precision of 0.89 on the training dataset and 0.97 on the test dataset, highlighting the potential of quantum-based methods for effectively detecting malicious URLs.

Ray et al.\cite{r7} developed a Quantum Machine Learning (QML) approach for improving phishing detection in the Ethereum transaction network. The proposed method utilizes Variational Quantum Classifier (VQC) and Quantum Support Vector Machine (QSVM) algorithms, applied to a dataset scraped from Etherscan, which contains both labeled phishing accounts and a significantly larger number of unlabeled legitimate accounts. To address the class imbalance, the authors focused on a training set of 300 data points and a test set of 300 data points, extracting seven statistical features such as in-degree, out-degree, and transaction strength. The data was encoded using a second-order Pauli-Z (ZZ) feature map and QRAC embedding. The VQC model, tested on a simulator with 32 qubits and using the ZZ feature map, achieved a macro F1 score of 0.67, phishing F1 score of 0.63, and phishing recall of 0.55. However, when run on quantum hardware, the model's performance was limited by longer training times, which required reducing the dataset size for practical implementation. Despite these challenges, the results demonstrate the potential of QML for detecting phishing in Ethereum networks.

Abreu et al.\cite{r11} showed that VQC achieves the highest F1 scores among QML models across datasets outperforming traditional ML methods. For instance, VQC scored 88.10\% on the UNSW-NB15 dataset, surpassing SVM's 82.34\%. Similarly, in the CIC-IDS-17 dataset, QCNN reached 95.60\%. While QML models excel in detecting specific attacks like Analysis, DDoS, and DoS traditional methods still outperform them in categories such as Worms and Shellcode. These results highlight the strengths of QML models but also the continued effectiveness of traditional methods suggesting the need for further research in QML for cybersecurity.

\section{Methodology}
\subsection{The Proposed Method}
The process begins with dataset preprocessing to scale features (e.g., feature1 to featureN) into
numerical values with

\begin{figure}[htbp]
    \centering
    \includegraphics[width=\linewidth]{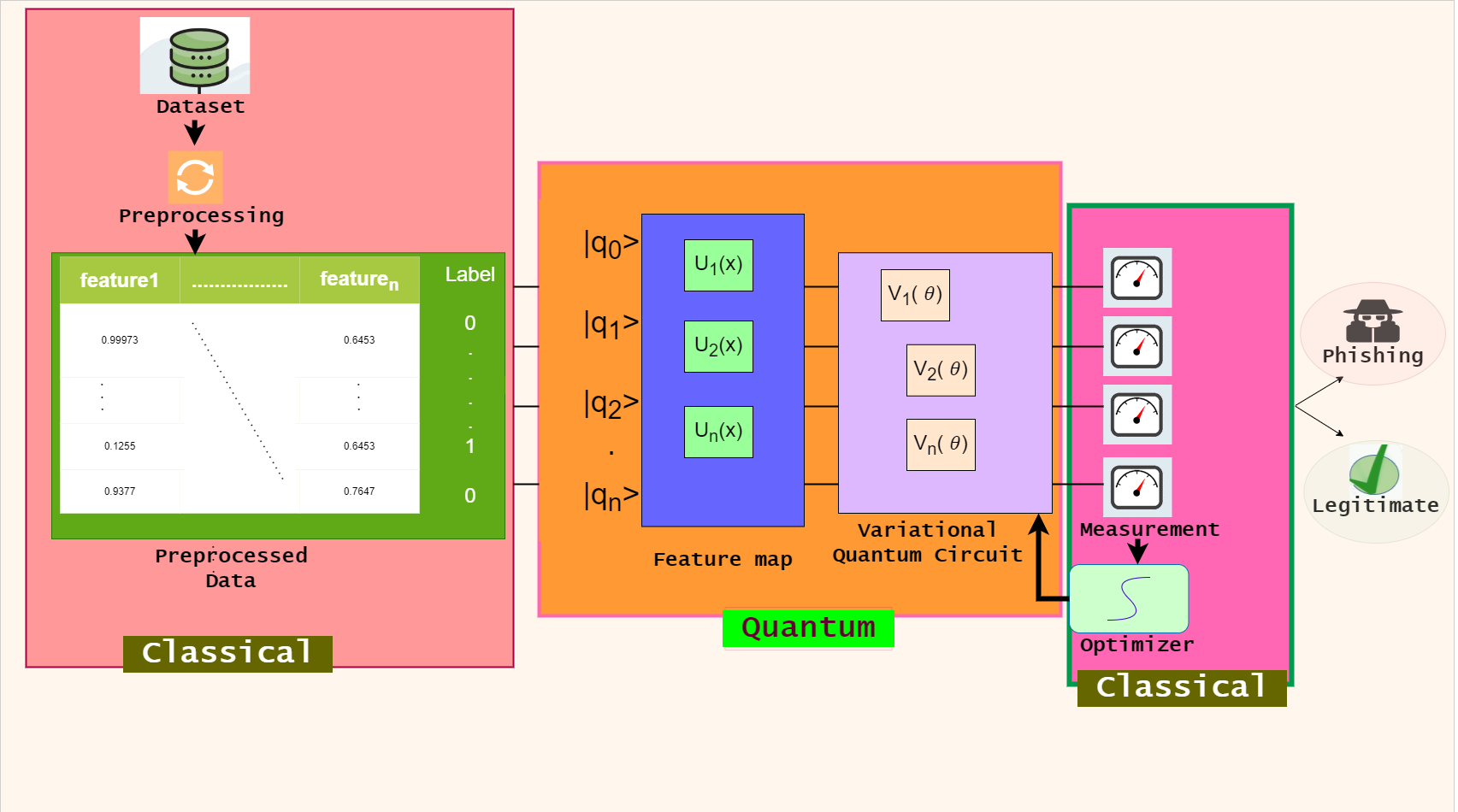}  
    \caption{The PhishVQC model for phishing detection.}
    \label{fig:model}
\end{figure}

associated 
labels (1 for legitimate, 0 for phishing). These features are encoded
into qubits 
\( (q_0, q_2, \dots, q_n) \)
using a quantum feature map. A variational quantum circuit (VQC) with adjustable parameters 
\( V(\theta) \) 
processes the qubits. Measurements

are taken and a classical optimizer updates the parameters to minimize error. The final measurement classifies outputs as phishing or legitimate.

\subsection{The URL Phishing Dataset}

The dataset used in this study is the PhiUSIIL Phishing URL Dataset \cite{r12} which contains 134,850 legitimate URLs and 100,945 phishing URLs. Sourced from the UCI Machine Learning Repository, the dataset includes 54 features that describe the 
structural and content-based characteristics of

 URLs.
Key features include the number of subdomains ('NoOfSubDomain'), the presence of obfuscation ('HasObfuscation'), the count of obfuscated characters ('NoOfObfuscatedChar'), the letter-to-character ratio in the URL ('LetterRatioInURL'), the number of question marks in the URL ('NoOfQMarkInURL'), and the number of lines of code in the URL ('LineOfCode').In the dataset, Class 1 consists of legitimate URLs while Class 0 includes phishing URLs. These features provide valuable insights for differentiating between legitimate and phishing URLs based on their structural patterns and content characteristics.

\subsection{Data Preprocessing}
Missing or null values in the dataset are checked. A good feature set consists of features that are strongly correlated with the target class but uncorrelated with each other . To reduce the complexity of quantum processing, the correlation between features is analyzed, and a threshold value of 0.5 is set to identify the most important features, reducing the feature count from 54 to 32.Finally, normalization is applied to prepare the classical data for quantum processing.

\subsection{Encoding Classical Data into Quantum States}
For quantum machine learning, classical data must be represented in a way that can be efficiently processed by quantum circuits.

\textit{Amplitude Encoding with Raw Feature Vector:}
In this context, we use amplitude encoding which requires that the number of features be a power of two. This is because the quantum state corresponding to the data must have a Hilbert space dimension that is a power of two, which is a fundamental requirement for amplitude encoding.

\begin{equation}
n = \log_2\left[\text{number of features}\right]
\end{equation}

Where n is the number of qubits required for encoding the features. The total dimension of the quantum state then becomes \( 2^n \) . If the number of features is not equal to \( 2^n \) after determining n, the number of features is padded with zeros to ensure the total number of features equals \( 2^n \). This enables us to encode the data into a quantum state using amplitude encoding.

To encode the classical data into quantum states, we use the Raw Feature Vector circuit. This circuit is a parameterized quantum circuit that initializes a quantum state based on the classical data. In the context of amplitude encoding, the classical data is mapped to the amplitudes of the quantum state.
The Raw Feature Vector circuit in Qiskit \cite{qiskit-rawfeaturevector} provides a direct, untransformed encoding of classical features into quantum states. The circuit works by applying an initialization operation, where the classical data defines the amplitudes of the quantum state.

 The RFV circuit requires a number of qubits equal to \( n \), where \( n \) is the logarithm of the feature dimension, \( \text{dim} \). Specifically, \( n = \log_2(\text{dim}) \). The quantum state is initialized accordingly.
The quantum state is initialized as follows:

\begin{equation}
|\psi\rangle = \sum_{i=0}^{2^n-1} c_i |i\rangle
\end{equation}
The quantum state \( |\psi\rangle \) is expressed as a linear combination of the computational basis states \( |i\rangle \) with corresponding amplitudes \( c_i \) derived from the input classical data.

This RawFeatureVector circuit directly initializes the quantum state based on the classical data, where the input data determines the amplitudes of the quantum state.
The Raw Feature Vector does not apply kernel transformations, which means it directly encodes the raw features into the quantum state. This is an essential part of amplitude encoding.

\subsection{Quantum Ansatz}

To perform quantum classification, a variational quantum circuit is used. This variational circuit is designed to approximate the target quantum state that represents the classification decision boundary. In this study, we use the RealAmplitude and EfficientSU2 ansätze which are parameterized quantum circuits capable of expressing complex quantum states. These ansatz also function as quantum feature maps. Since they are parameterized circuits, they can be considered ansatz in VQC.

\begin{figure}[htbp]
    \centering
    \includegraphics[width=\linewidth]{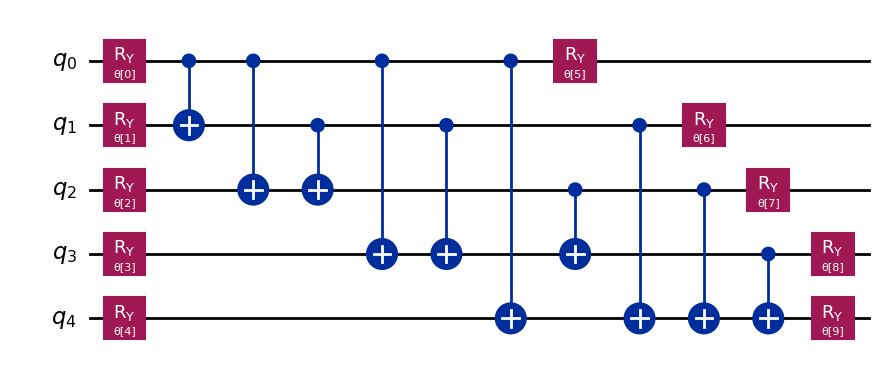}  
    \caption{The RealAmplitude Ansatz.}
    \label{fig:real}
\end{figure}

\begin{figure}[htbp]
    \centering
    \includegraphics[width=\linewidth]{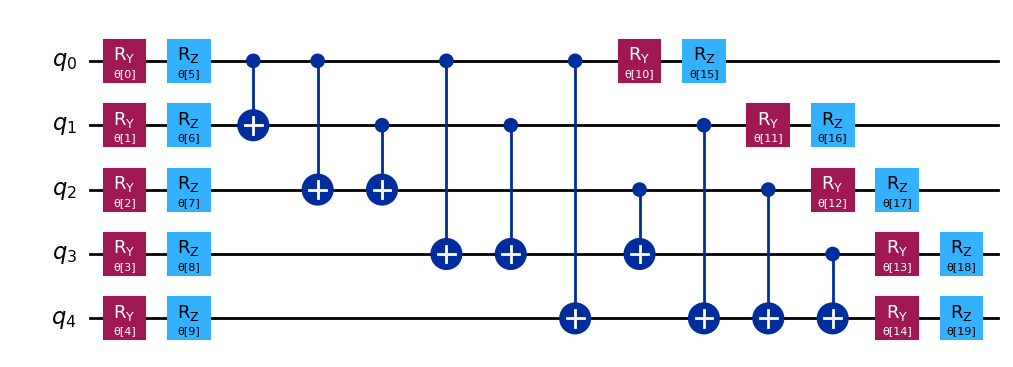}  
    \caption{The EfficientSU2 Ansatz.}
    \label{fig:su}
\end{figure}

\subsubsection{RealAmplitude Circuit: }The RealAmplitude ansatz \cite{qiskit-realamplitudes} typically includes entangling gates such as CNOT (controlled-NOT gates) that introduce correlations between qubits. These entangling gates are necessary to represent non-product states and are crucial for learning complex patterns in the data.

\subsubsection{EfficientSU2 Circuit: }
 The EfficientSU2 ansatz applies a sequence of parameterized rotations and entangling gates to prepare the quantum state. The rotation gates manipulate the quantum state, while the entangling gates (e.g., CNOT gates) create correlations between qubits.

The number of qubits is set to n, which corresponds to the number of features after zero padding.The ansatz can be repeated multiple times to increase the circuit depth and express more complex states. In this study, we use 3 and 4 repetitions. The entanglement='full' option means that all qubits are entangled in every layer allowing the quantum circuit to explore a broader state space and create more complex entanglements between qubits.

\subsection{Training the Variational Quantum Classifier (VQC)}

The Variational Quantum Classifier (VQC) combines the quantum feature map (RawFeatureVector) with the variational ansatz (RealAmplitude or EfficientSU2) to perform classification. The VQC is trained by adjusting the parameters of the ansatz to minimize the classification error.

\textit{Optimizer:} We use the COBYLA optimizer with a maximum of 300 iterations to minimize the objective function. COBYLA is a derivative-free optimization algorithm making it suitable for optimizing quantum circuits where gradients are not easily computed.

\textit{StatevectorSampler:} This sampler \cite{qiskit-statevectorsampler} is used to compute the state vector, which gives the full quantum state of the system after applying the quantum operations.

\textit{Callback:} A callback function is used to track the optimization progress by plotting the objective function value at each iteration. This helps to visualize how the optimization is proceeding.

\section{EXPERIMENTS AND RESULTS}

In Table \ref{tab:1} and Table \ref{tab:2}  ASetup and BSetup are two distinct configurations of the experimental environment for running the PhishVQC algorithm each with slightly different parameters like dataset sizes and feature map repetitions.The setup for the PhishVQC experiment involves two configurations: ASetup and BSetup, both utilizing a combination of classical and quantum libraries, including Qiskit, scikit-learn, pandas and others.In ASetup, the training dataset contains 640 samples with 160 samples for testing. The dataset has 32 features and the quantum model uses 5 qubits. The experiment runs for 300 iterations with the Feature Map encoding repeated 3 times. The quantum simulator used is the Statevector simulator.In BSetup, the dataset size increases to 960 training samples and 240 testing samples. The number of repetitions for the Feature Map encoding increases to 4, but other parameters remain unchanged. Both setups use the same quantum simulator and libraries.The key difference between the setups is the larger dataset and additional Feature Map repetition in BSetup aimed at examining the model's performance.

\begin{table}[htbp]
    \centering
    \caption{ASetup: The Setup for the first Experimental Environment of PhishVQC.}
    \includegraphics[width=\linewidth]{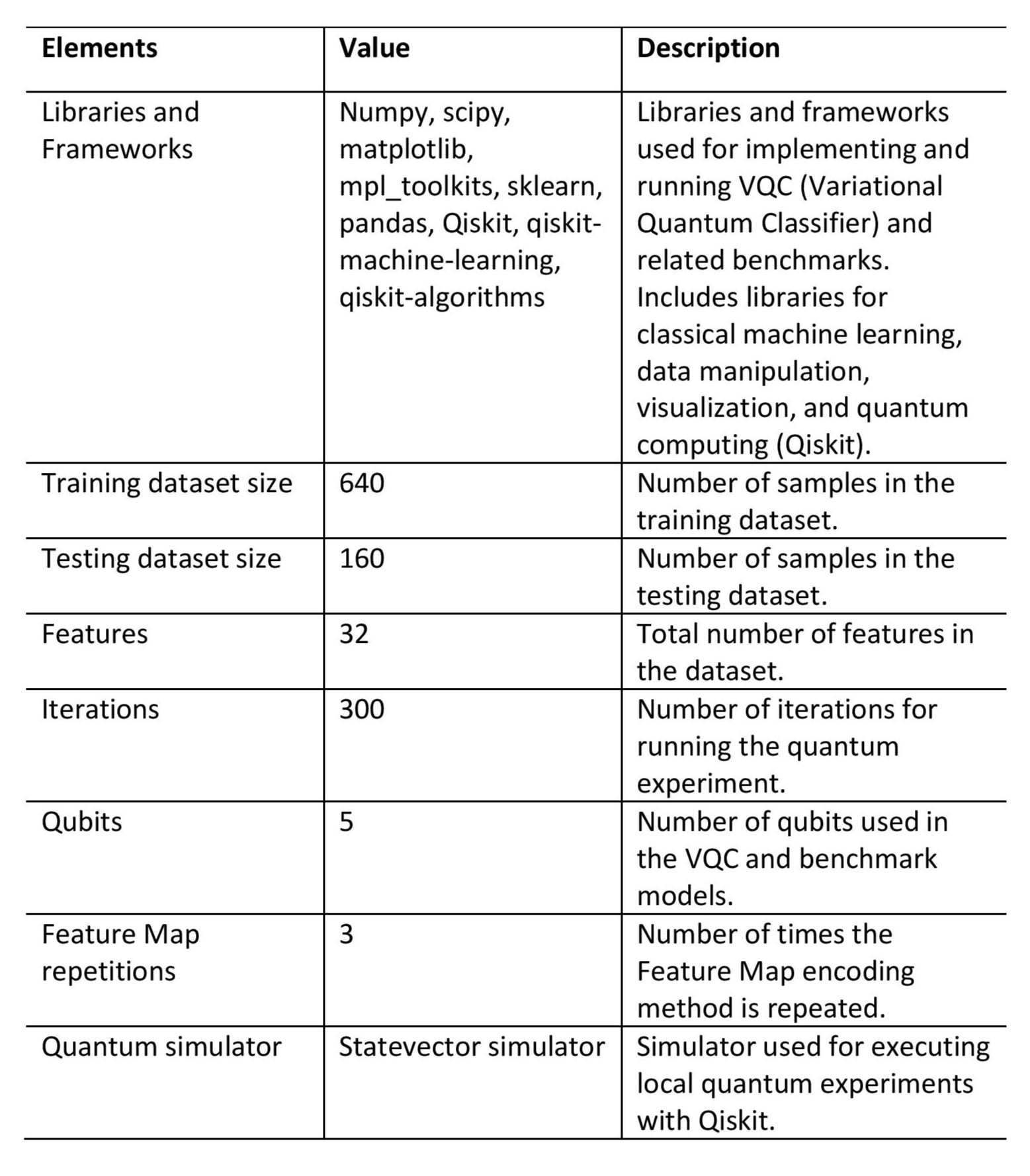}  
\label{tab:1}
\end{table}

\begin{table}[htbp]
    \centering
    \caption{BSetup: The Setup for the second Experimental Environment of PhishVQC.}
    \includegraphics[width=\linewidth]{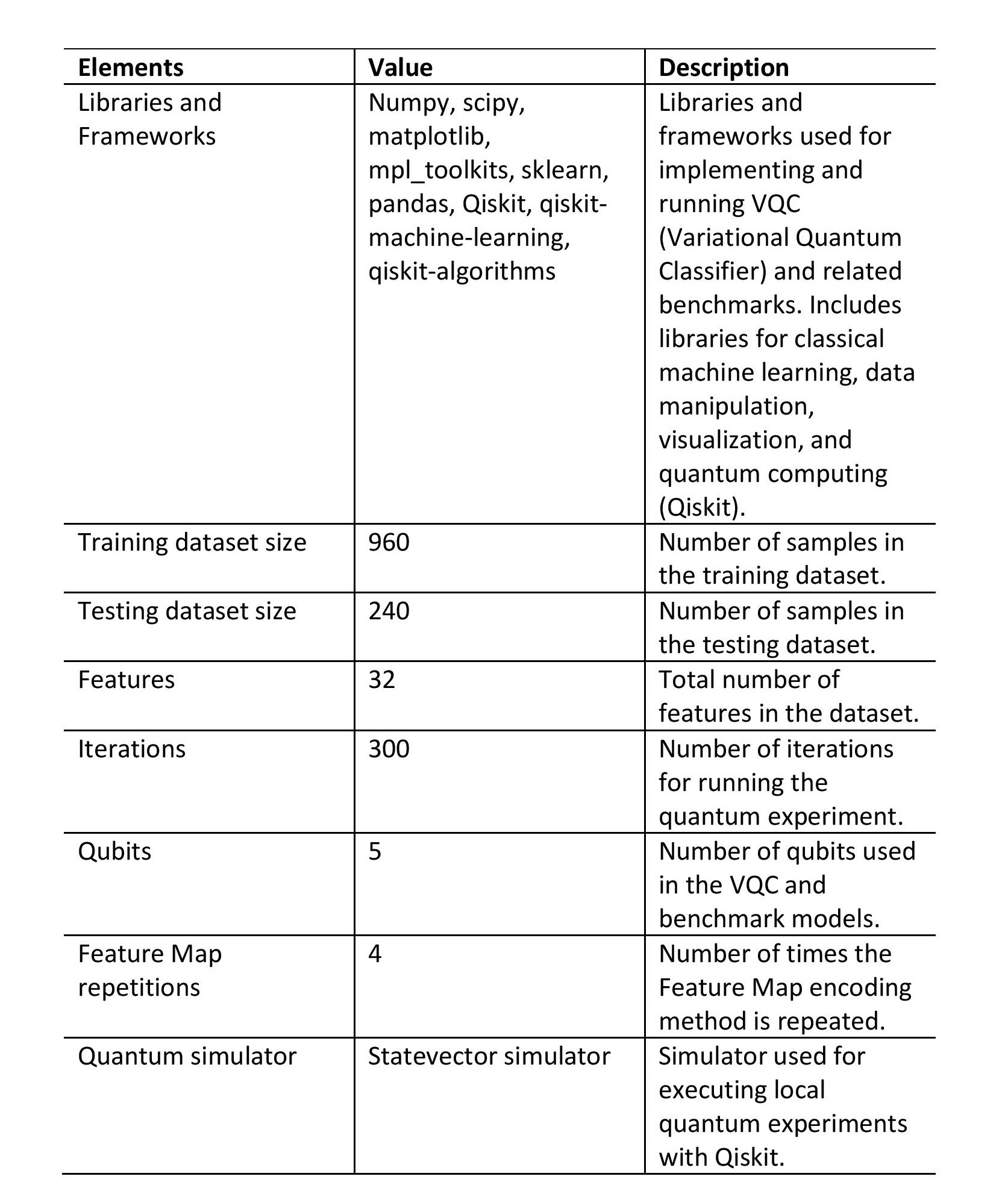}  
\label{tab:2}
\end{table}

\begin{table}
\centering
\caption{PhishVQC Performance with RealAmplitude and EfficientSU2 in ASetup and BSetup.}
\label{tab:3}
\resizebox{\columnwidth}{!}{
\begin{tabular}{|l|l|l|l|l|}
\hline
\textbf{Setup} & \textbf{Ansatz} & \textbf{Macro Phish Precision} & \textbf{Macro Phish Recall} & \textbf{Macro Avg F1-Score} \\ \hline
\multirow{2}{*}{ASetup} & RealAmplitude & 0.097 & 0.81 & 0.89 \\ \cline{2-5} 
& EfficientSU2 & 0.93 & 0.87 & 0.89 \\ \hline
\multirow{2}{*}{BSetup} & RealAmplitude & 0.96 & 0.80 & 0.88 \\ \cline{2-5} 
& EfficientSU2 & 0.90 & 0.84 & 0.87 \\ \hline
\end{tabular}
}
\end{table}

\begin{table}[!ht]
    \centering
    \caption{COMPARISON OF THE PhishVQC MODEL WITH Existing MODELS}
    \label{tab:4}
    \begin{tabular}{|c|c|}
    \hline
        \textbf{Model} & \textbf{Macro Avg F1-Score } \\ \hline
        VQC[\cite{r7}]  & 0.67 \\ \hline
        PhishVQC  & 0.89  \\ \hline
    \end{tabular}
\end{table}

In Table \ref{tab:3}   In the ASetup, Realamplitude demonstrates a strong performance with a phishing precision of 0.97 and a recall of 0.81 leading to a macro average F1-score of 0.89. This indicates that while Realamplitude excels at identifying phishing URLs with high precision, its ability to capture all phishing instances is slightly less effective, as shown by the recall value.EfficientSU2, on the other hand, shows a good balance with a phishing precision of 0.93 and a higher recall of 0.87 resulting in an equivalent macro average F1-score of 0.89. This highlights EfficientSU2's effectiveness in both identifying phishing URLs and covering a wider range of instances in the

\begin{figure}[htbp]
    \centering
    \includegraphics[width=\linewidth]{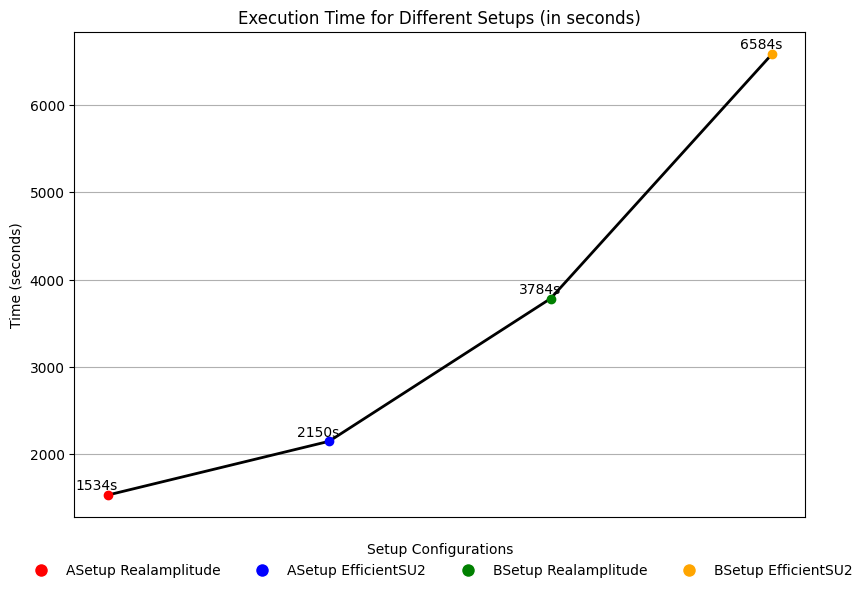}  
    \caption{Wall Time of Execution for PhishVQC with Different Setups.}
\label{fig:time}
\end{figure}

ASetup.

In the BSetup, Realamplitude's phishing precision remains high at 0.96 but its recall drops slightly to 0.80 leading to a macro F1-score of 0.88. This reduction in recall suggests that while the model maintains high precision, it is less effective at identifying all phishing instances in this setup.EfficientSU2

in BSetup shows a decline in phishing precision (0.90), but its recall improves to 0.84 resulting in a slightly lower F1-score of 0.87. This demonstrates that EfficientSU2's performance is still reliable but sees a small drop in phishing precision compared to ASetup.

Overall while both models appears with strong phishing detection capabilities, Realamplitude's performance is more consistent in terms of precision while EfficientSU2 shows better recall in the BSetup. However, the analysis of legitimate URLs is less significant in this context as the primary focus is on phishing URL detection though the macro scores in this case also remains 0.83 to 0.89.But falsely identifying a phishing URL as legitimate poses a significant security risk potentially allowing harmful websites to be accessed , the accurate detection of phishing URLs far more critical in ensuring security. In table \ref{tab:4} our  PhishVQC shows an improvement of 0.22 in the macro average F1-score  which indicates a 22\% increase in overall performance from previous study.

The figure \ref{fig:time} compares execution wall times for setups with varying data scales. As BSetup's datapoints increase by 50\% compared to ASetup, the wall time increases by almost three times. In ASetup, EfficientSU2 is about 40\% slower than Realamplitude (1534s). Similarly, BSetup EfficientSU2 (6584s) takes nearly twice as long as Realamplitude. This highlights that EfficientSU2 consistently requires more time than Realamplitude across both setups.

\section{DISCUSSIONS AND FUTURE WORK}

In this study the focus is placed on addressing the cybersecurity challenge of phishing URL detection in the context of quantum computing. The primary goal is to explore the potential of applying quantum machine learning techniques to improve phishing URL detection with a larger dataset and to identify an effective variational quantum classifier for URL classification.

PhishVQC shows that increasing the dataset size and feature map or ansatz repetitions in BSetup leads to minor improvements in performance but the key takeaway is that feature map repetitions have a more significant impact on the model's ability to capture phishing URLs. The larger dataset alone does not drastically change results. This highlights the importance of circuit complexity, especially through feature map repetitions over merely increasing dataset size for improving phishing detection accuracy.An increase in dataset size impacts the wall time execution of quantum circuits and the repetition of feature maps in the ansatz can also extend the execution time. Compared to the previous study  PhishVQC shows a 22\% improvement in macro F1-score.

In the transition from classical to quantum machine learning the current hardware complexity and challenges in processing large datasets make utilizing large sample sizes for such tasks still difficult. Nevertheless, this study highlights the essence of quantum machine learning (QML) for phishing URL detection which could inspire the exploration of other quantum algorithms such as Quantum Support Vector Machines (QSVM), Quantum Neural Networks (QNN) and Quantum Convolutional Neural Networks (QCNN) in the near future.







\bibliographystyle{ieeetr}
\bibliography{references}

\end{document}